\documentclass[prl,10pt,amsmath,twocolumn,floatfix]{revtex4}

\usepackage{amsmath}
\usepackage{amsfonts}
\usepackage{amssymb}
\usepackage{graphicx}
\usepackage{picture,xcolor}
\usepackage{color}
\usepackage[pdftex]{hyperref}  \hypersetup{bookmarks=true}

\begin{document}
\def\tr{{\rm tr}}
\def\sF{\mathcal{F}}

\title{Optimal quantum-enhanced interferometry using a laser power source}

\author{Matthias D.~Lang}
\email{mlang@unm.edu}
\affiliation{Center for Quantum Information and Control, University
of New Mexico, Albuquerque, NM 87131-0001, USA}
\author{Carlton M. Caves}
\email{ccaves@unm.edu}
\affiliation{Center for Quantum Information and Control, University of
New Mexico, Albuquerque, NM 87131-0001, USA}
\affiliation{Centre for Engineered Quantum Systems, School of
Mathematics and Physics, University of Queensland, Brisbane,
Queensland 4072, Australia}

\date{\today}

\begin{abstract}
We consider an interferometer powered by laser light (a coherent
state) into one input port and ask the following question: what is
the best state to inject into the second input port, given a
constraint on the mean number of photons this state can carry, in
order to optimize the interferometer's phase sensitivity?  This
question is the practical question for high-sensitivity
interferometry.  We answer the question by considering the quantum
Cram{\'e}r-Rao bound for such a setup.  The answer is squeezed
vacuum.
\end{abstract}

\maketitle

The discovery that squeezed vacuum, injected into the normally unused
port of an interferometer, provides phase sensitivity below the
shot-noise limit~\cite{Caves1981a} led to thirty years of technology
development, beginning with initial proof-of-principle
experiments~\cite{MXiao1987a,Grangier1987a} and culminating recently
in the use of squeezed light to beat the shot-noise limit in the
GEO~600 gravitational-wave detector~\cite{LIGO2011a} and the
Hanford LIGO detector~\cite{LIGO2013a}.

In the last decade much work has been devoted to exploring ultimate
quantum limits on estimating the differential phase shift between two
optical paths and to finding the states that achieve these limits.
Given exactly $N$ photons, the optimal state, in the absence of
photon loss, is a \textit{N00N} state,
$(|N,0\rangle+|0,N\rangle)/\sqrt2$~\cite{Gerry2000a,Boto2000a,Gerry2002a,Lee2002a},
i.e., a superposition of all photons proceeding down one path with
all photons proceeding down the other path.  The \textit{N00N} state is the
optical analogue of the cat state that is optimal for atomic (Ramsey)
interferometry~\cite{Bollinger1996a}.  Since the \textit{N00N} state is
extremely sensitive to photon loss, considerable effort has gone into
determining optimal $N$-photon input states and corresponding
sensitivities in the presence of photon
loss~\cite{Dorner2009a,Knysh2011a,Escher2011a,Escher2011b}.

While these states indeed deliver optimal or near-optimal performance,
given a fixed input energy, we argue that they are not of practical
relevance because they are very hard to produce with current
technology and are therefore only available with quite low photon
numbers.  Consequently, the phase resolution obtained from using
these optimal states cannot compete with the resolution obtained from
a classical interferometer operating at or near the shot-noise limit
with a strong, commercially available laser.

This does not mean, however, that nonclassical states are useless for
metrology.  The use of squeezed states to enhance the sensitivity of
the GEO~600 and LIGO interferometers is testimony to the efficacy of
squeezed light in a situation where the lasers powering the
interferometer have been made as powerful as design constraints
allow.  This paper turns the focus away from states that have
only been created with very small numbers of photons and instead
investigates a particular, practical question: when an interferometer
is powered by a laser producing coherent-state light, what is the
best state to put into an interferometer's secondary input port?
The answer is not surprising: squeezed vacuum.

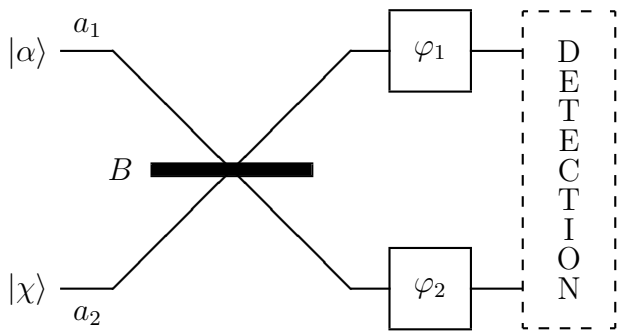
\begin{figure}[h]
\begin{center}
\begin{picture}(240,130)
\thicklines

\put(0,106){\large$|\alpha \rangle$}
\put(0,17){\large$|\chi\rangle$}

\put(38,61){\large$B$}
\put(25,117){\large$a_1$}
\put(20,110){\line(1,0){20}}
\put(20,20){\line(1,0){20}}
\put(25,8){\large$a_2$}
\put(40,20){\line(1,1){90}}
\put(40,110){\line(1,-1){90}}
\setlength\fboxsep{0pt}
\put(55,63){\colorbox{black}{\framebox(60,4){}}}

\put(130,20){\line(1,0){15}}
\put(130,110){\line(1,0){15}}

\put(145,5){\framebox(30,30){\large$\varphi_2$}}
\put(145,95){\framebox(30,30){\large$\varphi_1$}}

\put(175,20){\line(1,0){20}}
\put(175,110){\line(1,0){20}}

\put(195,5){\dashbox{4}(35, 120){\shortstack{\large D\\\large E\\\large T\\\large E\\\large C\\\large T\\\large I\\\large O\\\large N}}}
\end{picture}
\caption{Measurement of a differential phase shift.  An (upper) mode
$a_1$ in a coherent state $|\alpha\rangle$ and a (lower) mode $a_2$
in an arbitrary pure state $|\chi\rangle$ are incident on a 50:50
beam splitter, which performs the unitary transformation~$B$ of
Eq.~(\protect{\ref{eq:BS}}).  After the beam splitter, phase shifts
$\varphi_1$ and $\varphi_2$ are imposed in the two arms; the action
of the phase shifters is contained in the unitary operator $U$ of
Eq.~(\protect{\ref{eq:U}}). Finally, a measurement is made to detect
the phase shifts.  When the measurement is pushed beyond a second
50:50 beam splitter, the result is a Mach-Zehnder interferometer, which
is sensitive only to the differential phase shift
$\phi_d=\varphi_1-\varphi_2$.} \label{setting}
\end{center}
\end{figure}

The setting for our analysis is depicted in Fig.~\ref{setting}.
Specifically, we consider a situation where laser light, described by
a coherent state $|\alpha\rangle=D(a_1,\alpha)|0\rangle$ of a mode
$a_1$, is fed into the primary input port of a 50:50 beam splitter.
The secondary input port is illuminated by mode $a_2$, which is in an
arbitrary pure state $|\chi\rangle$.  The beam splitter performs the
unitary transformation
\begin{align}
B=e^{-iJ\pi/4}\;,\qquad
J\equiv a_1^{\dagger}a_2+a_2^{\dagger}a_1\;.
\label{eq:BS}
\end{align}
The two optical paths after the beam splitter experience phase shifts
$\varphi_1$ and $\varphi_2$; the phase-shift unitary operator is
\begin{align}
U
=e^{i(\varphi_1\,a_1^\dagger a_1+\varphi_2\,a_2^\dagger a_2)}
=e^{iN_s\phi_s/2}e^{iN_d\phi_d/2}\;.
\label{eq:U}
\end{align}
In the second form we introduce the sum and difference phase shifts,
$\phi_s=\varphi_1+\varphi_2$ and $\phi_d=\varphi_1-\varphi_2$, and
the corresponding sum and difference number operators;
$N_s=a_1^\dagger a_1+a_2^\dagger a_2$ is the total number operator
for the two modes, and $N_d=a_1^\dagger a_1-a_2^\dagger a_2$ is the
number-difference operator.  We assume that there are no losses in
this configuration.  The two-mode state after the phase shifters~is
\begin{align}
|\psi\rangle=UB|\psi_{\rm in}\rangle\,,
\end{align}
where $|\psi_{\rm in}\rangle$ is the state before the beam splitter.

We use the quantum Fisher information (QFI) to investigate the
optimal resolution for estimating the phase shifts $\phi_s$ and
$\phi_d$.  The advantage of QFI is that it gives a bound on phase
resolution, called the quantum Cram{\'e}r-Rao bound (QCRB), that
applies to all quantum measurements on the two optical paths and to
all procedures for estimating the phase shifts from the measurement
results.  In particular, let $\phi_s^{\rm est}$ and $\phi_d^{\rm
est}$ denote estimators of the sum and difference phase shifts, and
introduce the covariance matrix of the estimators,
\begin{align}
\Sigma=
\begin{pmatrix}
\vphantom{\Big(}\langle(\Delta\phi_s^{\rm est})^2\rangle
&\langle\Delta\phi_s^{\rm est}\Delta\phi_d^{\rm est}\rangle\\
\langle\Delta\phi_d^{\rm est}\Delta\phi_s^{\rm est}\rangle
&\vphantom{\Big(}\langle(\Delta\phi_d^{\rm est})^2\rangle
\end{pmatrix}\;,
\end{align}
where here and throughout $\Delta O\equiv O-\langle O\rangle$ denotes
the deviation of a quantity from its mean.  The QCRB is the matrix
inequality
\begin{align}
\Sigma\ge\sF^{-1}\,,
\label{eq:QCRB}
\end{align}
where $\sF$ is the (real, symmetric) Fisher-information
matrix~\cite{Helstrom,Holevo}.  The matrix QCRB implies that
$\tr\Sigma\ge\tr\sF^{-1}$ and $\det\Sigma\ge\det\sF^{-1}$; for more
than one parameter, the matrix QCRB cannot generally be
saturated~\cite{Fujiwara2006}.

For pure states, the Fisher-information matrix is given
by~\cite{Jarzyna2012a}
\begin{align}
\sF_{\!jk}
= 4\,\bigl(\langle \partial_j \psi | \partial_k \psi\rangle
-\langle \partial_j \psi|\psi\rangle \langle \psi | \partial_k \psi\rangle\bigr)\;,
\end{align}
where $j$ and $k$ take on the values $s$ and $d$ and thus the
derivatives are with respect to $\phi_s$ and $\phi_d$.  We retain
both $\phi_s$ and $\phi_d$ in our analysis for the present, but
eventually specialize to estimation of the differential phase shift
alone.  This would be the case if the final measurement were moved
behind a second 50:50 beam splitter, giving a standard (Mach-Zehnder)
interferometric configuration.

There are important practical reasons for considering the
configuration of Fig.~\ref{setting}.  The first is that in a typical
phase measurement, the easiest way to improve sensitivity is to buy
more photons. The cheapest coherent source being a laser, the
relevant model is that of a laser producing an input coherent state
with the largest possible amplitude.  To avoid the phase noise of the
laser, either intrinsic or excess, one splits the laser light at a
50:50 beam splitter.  Phase shifts are imposed in the two arms, and
then in a Mach-Zehnder configuration, the light in the two arms is
recombined at a second 50:50 beam splitter, after which differenced
photodetection or differenced homodyne detection is used to detect
the differential phase shift.  This interferometric technique is
insensitive to the common-mode phase shift $\phi_s$ in the two arms,
which is just another way of saying that it is insensitive to the
laser noise.  Yet another way of putting this is that each arm serves
as a phase reference for the other.

The Mach-Zehnder interferometric configuration gives
shot-noise-limited sensitivity when the secondary port is illuminated
by vacuum.  To go beyond the shot-noise limit, one must replace
the vacuum coming into the secondary port with some other,
nonclassical quantum state of light; this inevitably makes the light
in the two arms of the interferometer entangled, this modal entanglement
having been made by the input beam splitter~\cite{entanglement}.  A
major advantage of the setting in Fig.~\ref{setting} is that the main
power production is separated from the generation of nonclassical
light, which only has to get a phase reference from the laser.  Many
analyses of phase sensitivity start by asking what entangled state in
the two arms gives the best sensitivity, but this approach generally
requires an entangled state that cannot be made by beamsplitting a
product state and thus gives up the practical advantage of separating
the main power production from the production of nonclassical light.

In accordance with this discussion, the intended mode of operation of
our interferometer is to have the coherent state carry many more
photons than the light input to the secondary port.
Since it does not hinder our analysis, however, we allow for the opposite
possibility and all intermediate ones in our analysis.

An analysis similar in spirit to ours has investigated the best
performance of an interferometer, given a constraint on the total
mean number of photons, when the primary input port is illuminated
with many more photons than the secondary input port~\cite{DD2013}.
The results show that a coherent state input to the primary port and
squeezed vacuum into the other port comes very close to achieving a
bound on the Fisher information that applies to all input states,
both product and nonproduct states.  This result holds when the
photon loss exceeds a certain level, given in terms of the total mean
number of photons, and thus is complementary to our result.

In our setting, the Fisher matrix for an arbitrary input state
$|\psi_{\rm in}\rangle$ becomes
\begin{align}
\sF_{ss}
&=\langle\psi_{\rm in}|B^\dagger N_s^2 B|\psi_{\rm in}\rangle
-\langle\psi_{\rm in}|B^\dagger N_sB|\psi_{\rm in}\rangle^2\, ,\\
\sF_{dd}
&=\langle\psi_{\rm in}|B^\dagger N_d^2 B|\psi_{\rm in}\rangle
-\langle\psi_{\rm in}|B^\dagger N_d B|\psi_{\rm in}\rangle^2\, ,
\label{eq:Fdd}\\
\sF_{sd}&=\sF_{ds}
=\langle\psi_{\rm in}|B^\dagger N_sN_d B|\psi_{\rm in}\rangle\notag\\
&\phantom{=\sF_{ds}=}-\langle\psi_{\rm in}|B^\dagger N_sB|\psi_{\rm in}\rangle
\langle\psi_{\rm in}|B^\dagger N_d B|\psi_{\rm in}\rangle \, .
\end{align}
We can use $B^\dagger a_1B=(a_1-ia_2)/\sqrt 2$ and $B^\dagger
a_2B=(a_2-ia_1)/\sqrt 2$ to get $B^\dagger N_sB=N_s$ and
\begin{align}
B^\dagger N_d B=-i(a_1^\dagger a_2-a_2^\dagger a_1)\equiv K\;.
\end{align}
The Fisher matrix is thus the covariance matrix of $N_s$ and $K$ with
respect to the initial state.  Notice that $J_z=N_d/2$, $J_x=J/2$,
and $J_y=K/2$ make up the three components of an angular momentum and
provide a convenient way of analyzing
interferometry~\cite{Yurke1986a}.

For the product input that is our main concern,
\begin{align}
|\psi_{\rm in}\rangle=|\alpha\rangle\otimes|\chi\rangle,
\label{eq:psiin}
\end{align}
the Fisher matrix becomes, with $N_2=a_2^\dagger a_2$,
\begin{align}
\sF_{ss}
&=|\alpha|^2+\langle\chi|(\Delta N_2)^2|\chi\rangle\;,\\
\sF_{dd}
&=|\alpha|^2\langle \chi|
(\Delta a_2\Delta a_2^\dagger +\Delta a_2^\dagger\Delta a_2)|\chi\rangle\notag\\
&\phantom{=|\alpha|^2}
-\alpha^{*2}\langle\chi|(\Delta a_2)^2|\chi\rangle
-\alpha^2\langle\chi|(\Delta a_2^\dagger)^2|\chi\rangle\\
&\phantom{=|\alpha|^2}+\langle\chi|N_2|\chi\rangle\;, \notag\\
\sF_{sd}&=\sF_{ds}
=-i\alpha^*\langle\chi|N_2(\Delta a_2)|\chi\rangle
+i\alpha\langle\chi|(\Delta a_2^\dagger)N_2|\chi\rangle \notag\\
&\phantom{=\sF_{ds}=|\alpha|^2}-i\alpha^*\langle\chi|a_2|\chi\rangle\;.
\end{align}

Partly because the matrix QCRB cannot generally be saturated
\cite{Fujiwara2006}, but chiefly because we are mainly interested in
measurements of the differential phase shift, we specialize now to
single-parameter estimation of $\phi_d$, for which the QCRB reduces
to
\begin{equation}
\langle(\Delta\phi_d^{\rm est})^2\rangle=
\Sigma_{dd}\ge\frac{1}{\sF_{dd}}\;.
\end{equation}
It is known that there is a quantum measurement that achieves the
single-parameter QCRB~\cite{Braunstein1994a}, i.e., has the required
Fisher information, and it is also known that the resulting QCRB can
be attained asymptotically in many trials by maximum-likelihood
estimation.

What we do now is to maximize $\sF_{dd}$ over all initial states
$|\chi\rangle$ of mode~$a_2$ subject to a constraint of fixed mean
photon number $\bar N=\langle\chi|N_2|\chi\rangle$.  The optimal
state turns out to be squeezed vacuum with the requisite mean photon
number.  We then use results of Pezz{\'e} and
Smerzi~\cite{Pezze2008a} to indicate how the ultimate sensitivity can
be achieved in a Mach-Zehnder interferometer in which one does direct
photon detection of the two outputs.

To get started on maximizing $\sF_{dd}$, we assume, without loss of
generality, that $\alpha$ is real, and we write $\sF_{dd}$ in terms
of moments of the (Hermitian) quadrature components, $x$ and $p$, of
$a_2=(x+ip)/\sqrt2$:
\begin{align}
\sF_{dd}=2\alpha^2\langle(\Delta p)^2\rangle+\bar N\;.
\label{eq:Fdd}
\end{align}
Here and for the remainder of the paper, all expectation values are
taken with respect to the initial state~(\ref{eq:psiin}).  The first
term in Eq.~(\ref{eq:Fdd}), $2\alpha^2\langle(\Delta p)^2\rangle$, is
due to interference between the coherent state and the phase
quadrature $p$ of the light coming into the secondary port; if
$\alpha^2\gg\bar N$, this term dominates and gives the shot-noise
limit when mode~$a_2$ is in vacuum and improvements beyond shot noise
when $\langle(\Delta p)^2\rangle>1/2$.  If $\alpha=0$, the
contribution from $\bar N$ in Eq.~(\ref{eq:Fdd}) dominates and
expresses the shot-noise limit for illumination only through the
secondary~port.

We now maximize the variance of $p$, subject to a constraint on the
mean number of photons.  Writing
\begin{align}
2\bar N+1&=\langle p^2\rangle+\langle x^2\rangle\nonumber\\
&=\langle p\rangle^2+\langle x\rangle^2
+\langle(\Delta p)^2\rangle+\langle(\Delta x)^2\rangle\;,
\end{align}
we see that
\begin{equation}
\langle(\Delta p)^2\rangle+\langle(\Delta x)^2\rangle\le2\bar N+1\;,
\label{eq:xpboundone}
\end{equation}
with equality if and only if $\langle x\rangle=\langle p\rangle=0$.
We also have
\begin{align}
\bigl(\langle(\Delta p)^2\rangle&-\langle(\Delta x)^2\rangle\bigr)^2\nonumber\\
&=\bigl(\langle(\Delta p)^2\rangle+\langle(\Delta x)^2\rangle\bigr)^2
-4\langle(\Delta x)^2\rangle\langle(\Delta p)^2\rangle\nonumber\\
&\le-1+\bigl(\langle(\Delta p)^2\rangle+\langle(\Delta x)^2\rangle\bigr)^2\nonumber\\
&\le4\bar N(\bar N+1)\;,
\label{eq:xpboundtwo}
\end{align}
with equality in the first inequality if and only if $|\chi\rangle$
is a minimum-uncertainty state, i.e., $\langle(\Delta
x)^2\rangle\langle(\Delta p)^2\rangle=1/4$.  Combining
Eqs.~(\ref{eq:xpboundone}) and~(\ref{eq:xpboundtwo}) bounds
$\langle(\Delta p)^2\rangle$ and hence gives a bound on the Fisher
information,
\begin{equation}
\sF_{dd}
\le\alpha^2\!\left(2\bar N+2\sqrt{\bar N(\bar N+1)}+1\right)
+\bar N\equiv\sF_{\rm max}\;,
\label{eq:Fmax}
\end{equation}
with equality if and only if $|\chi\rangle$ is a zero-mean
minimum-uncertainty state, i.e., the squeezed vacuum state
$e^{r(a^2-a^{\dagger 2})/2}|0\rangle$, with $\bar N=\sinh^2\!r$.  In
terms of the squeeze parameter $r$, the bound on the Fisher
information takes the simple form $\sF_{\rm
max}=\alpha^2e^{2r}+\sinh^2\!r$.

It is useful to manipulate the bound~(\ref{eq:Fmax}) in the following
way:
\begin{align}
\sF_{\rm max}=4\alpha^2\bar N+R=N_{\rm tot}^2-(\alpha^2-\bar N)^2+R\;\label{fisher}.
\end{align}
Here $N_{\rm tot}=\alpha^2+\bar N$ is the total mean number of
photons into both input ports, and the remainder term is given~by
\begin{equation}
R=\bar N+\alpha^2\!\left(2\sqrt{\bar N(\bar N+1)}-2\bar N+1\right)\;.
\end{equation}
Applying $\bar N\le\sqrt{\bar N(\bar N+1)}\le \bar N+\frac{1}{2}$, we
have $N_{\rm tot}=\alpha^2+\bar N\le R\le2\alpha^2+\bar N=N_{\rm
tot}+\alpha^2$.  When $N_{\rm tot}$ is large, the remainder term is
negligible compared to $N_{\rm tot}^2$.  Moreover, when
$\alpha^2=\bar N$, we have $\sF_{\rm max}=N_{\rm tot}^2+R$, which
gives the Heisenberg limit on phase sensitivity plus a small
correction that satisfies $N_{\rm tot}\le R\le 3N_{\rm tot}/2$.  The
apparent violation of the Heisenberg limit comes from not having a
fixed total number of photons.  That this configuration using
coherent and squeezed light can achieve the Heisenberg limit was
shown in \cite{Pezze2008a}.

The case of primary practical interest has $\alpha^2\gg\bar
N=\sinh^2\!r$, in which case the maximal Fisher information reduces
to $\sF_{\rm max}=\alpha^2 e^{2r}$.  This corresponds to the standard
picture of reduced fluctuations in the quadrature that produces
differential phase fluctuations in the interferometer, and it gives
the standard phase sensitivity, $1/\sqrt{\sF_{\rm
max}}=e^{-r}/\alpha$, for a squeezed-state interferometer.  Indeed,
the Fisher bound can be achieved by recombining the two optical paths
at a second 50:50 beam splitter to create an interferometer and
performing direct detection of the two outputs. The estimator can be
taken to be the standard linear estimator that inverts the fringe
pattern of the differenced photocount to estimate the differential
phase shift.

Though it might be surprising, squeezed vacuum remains the optimal
state into the secondary port even when the secondary port is allowed
as many or more photons as the coherent-state input.  We can appeal
to the results of Pezz{\'e} and Smerzi~\cite{Pezze2008a} to show that
the Mach-Zehnder configuration, with coherent-state and
squeezed-vacuum inputs and direct detection at the output, can
achieve the QCRB~(\ref{eq:Fmax}) for all values of the ratio
$\alpha^2/\bar N$.  Pezz{\'e} and Smerzi showed that for this
configuration, the \emph{classical\/} Fisher information of the
probability for the output sum and and difference photocounts,
$P(n_s,n_d|\phi_d)=P(n_d|n_s,\phi_d)P(n_s)$, is equal to $\sF_{\rm
max}$~\cite{suppmat}.  When $\alpha^2\alt\bar N=\sinh^2r$, however,
the interferometer is running partially or mainly on the phase
dependence of the squeezed vacuum noise, and the standard linear
estimator mentioned above does not deliver optimal
sensitivity~\cite{Pezze2008a,Bondurant1984a}.  Indeed, one can use
the convexity of the Fisher information~\cite{MCohen1968a} to show
that any estimator that uses only the differenced photocount $n_d$,
ignoring the sum photocount $n_s$, does worse than keeping
both~\cite{footnote}, even though $n_s$ is insensitive to the
differential phase shift $\phi_d$. Instead of using an estimator to
verify that the classical Fisher bound---and, hence, from our
analysis, the QCRB---can be achieved, Pezz{\'e} and Smerzi simulated
a Bayesian analysis that indicates the classical bound can be
achieved for all ratios $\alpha^2/\bar N$.

We note that squeezed vacuum is not the state that maximizes the
entanglement of the two optical paths after the input beam splitter.
A number state $|\bar N\rangle$ in the second mode leads to a larger
value of the marginal entropy of the two paths~\cite{Asboth2005a}.

We have analyzed a real-world scheme for measuring differential phase
shifts, in which a coherent state illuminates one side of a 50:50
beam splitter and an arbitrary quantum state of light the other.  We
showed that given a constraint on the total mean number of photons,
the optimal state to put into the secondary input port is a squeezed
vacuum state, regardless of the relative mean photon numbers of the
two inputs.  At least two questions beg for further attention. We do
not know a simple optimal estimator when the squeezed light carries
as many or more photons than the coherent input.  We assumed no
photon losses throughout our analysis and thus do not know the
optimal state to put into the secondary port in the presence of
losses, even though we suspect---and the results of
Ref.~\cite{DD2013} suggest---that it is squeezed vacuum.

MDL thanks S.~Pandey for useful discussions.  This work was supported
in part by National Science Foundation Grants No.~PHY-1212445 and
PHY-1005540 and by Office of Naval Research Grant
No.~N00014-11-1-0082.

\clearpage
\begin{widetext}

\section{Supplementary Material}
\subsection{Classical Fisher Information for an interferometric configuration}

In this section we guide the reader through the calculation of
classical Fisher information for a Mach-Zehnder interferometer with
direct detection at the output and show that it is the same as the
quantum Fisher information for a broad class of input states.  The
result of this calculation for the coherent-state--sqeezed-vacuum
input was reported by Pezz\'{e} and Smerzi~\cite{Pezze2008a}.

By adjusting phases at the second 50:50 beam splitter in the
Mach-Zehnder, we can let it be described by the unitary
operator~$B^\dagger$.  With this choice the Mach-Zehnder performs the
overall transformation $B^\dagger
UB=e^{iN_s\phi_s/2}e^{iK\phi_d/2}=e^{iN_s\phi_s/2}e^{iJ_y\phi_d}$,
which includes the common-mode phase shift $\phi_s$ and a $J_y$
rotation of the input state by angle $-\phi_d$~\cite{Yurke1986a}.
The angle $\phi_d$ is the relative phase shift in the two arms, which
we are trying to estimate.  The probability for detecting $n_1$
photons in the first output mode $a_1$ and $n_2$ photons in the
second output mode $a_2$ is
\begin{align}
P(n_1,n_2|\phi_d)
=|\langle n_1,n_2|e^{iN_s\phi_s/2}e^{iJ_y\phi_d}|\psi_{\rm in}\rangle|^2
=|\langle n_1,n_2|e^{iJ_y\phi_d}|\psi_{\rm in}\rangle|^2\,.
\end{align}
This is also the probability to detect $n_s=n_1+n_2$ total photons at
the output and a difference $n_d=n_1-n_2$; written in terms of the
eigenstates of sum and difference photon numbers, this probability
becomes
\begin{align}
P(n_s,n_d|\phi_d)=|\langle n_s,n_d|e^{iJ_y\phi_d}|\psi_{\rm in}\rangle|^2\,.
\end{align}
It is convenient to switch to the angular-momentum basis
$|j,m\rangle=|n_s,n_d\rangle$ by identifying $j=n_s/2$ and $m=n_d/2$.
The probability becomes
\begin{align}
P(n_s,n_d|\phi_d)
=|\langle j,m|e^{iJ_y\phi_d}|\psi_{\rm in}\rangle|^2
=\Biggl|\sum_{m'=-j}^j d^j_{m,m'}(-\phi_d)\langle j,m'|\psi_{\rm in}\rangle\Biggr|^2\,,
\label{eq:Pnsnd}
\end{align}
where we use the fact that $J_y$ conserves total photon number (total
angular momentum) and we introduce the Wigner rotation matrices,
\begin{align}
d^j_{m,m'}(-\phi_d)
=\langle j,m|e^{i\phi_d J_y}|j,m'\rangle
=\langle j,m'|e^{-i\phi_d J_y}|j,m\rangle\,.
\label{eq:djm}
\end{align}
Equation~(\ref{eq:Pnsnd}) is the form of the joint probability quoted
in Eq.~(5) of Ref.~\cite{Pezze2008a}.

The case of interest is a product input, $|\psi_{\rm
in}\rangle=|\alpha\rangle\otimes S|0\rangle$, where
$S|0\rangle=|0\rangle$ is the optimal squeezed vacuum input to the
secondary input port.  Without loss of generality, we can assume that
$\alpha$ is real.  Then the optimal squeezed vacuum state is squeezed
along the quadrature axes of the input mode $a_2$, and the input
state takes the form
\begin{align}
|\psi_{\rm in}\rangle=|\alpha\rangle\otimes S(r)|0\rangle\,,\qquad
S(r)=e^{r(a^2-a^{\dagger 2})/2}\,.
\label{eq:alphaS0}
\end{align}
Under these assumptions, the amplitude $\langle j,m|\psi_{\rm
in}\rangle=\langle n_s,n_d|\psi_{\rm in}\rangle$ is real.  That these
amplitudes are real is the only assumption about the input state that
we use in the following; our calculation thus applies to all product
and nonproduct inputs for which these amplitudes are real. Since the
Wigner rotation matrix is real, as displayed in Eq.~(\ref{eq:djm}),
the sum in Eq.~(\ref{eq:Pnsnd}) is also real.

The classical Fisher information for a Mach-Zehnder interferometer
with direct detection is
\begin{align}
F(\phi_d)
=\sum_{n_s,n_d} \frac{1}{P(n_s,n_d|\phi_d)}
\left(\frac{\partial P(n_s,n_d|\phi_d)}{\partial \phi_d}\right)^2\,.
\label{fishdef}
\end{align}
Omitting indices and decorations for readability when there is no
risk of confusion, we can write
\begin{align}
\frac{\partial P}{\partial \phi_d}
=2\sqrt{P}
\sum_{m'=-j}^j\langle j,m'|\psi_{\rm in}\rangle
\frac{\partial d^j_{m,m'}(-\phi_d)}{\partial\phi_d}\,,
\end{align}
where we use the reality of the number-state expansion coefficients
of $|\psi_{\rm in}\rangle$, and thus
\begin{align}
F(\phi_d)
&=4\sum_j\sum_{m=-j}^j
\Biggl(\sum_{m'=-j}^j\langle j,m'|\psi_{\rm in}\rangle
\frac{\partial d^j_{m,m'}(-\phi_d)}{\partial\phi_d}\Biggr)^2\nonumber\\
&=4\sum_j\sum_{m=-j}^j\sum_{q,q'=-j}^j
\langle\psi_{\rm in}|j,q\rangle\langle j,q|J_ye^{-iJ_y\phi_d}|j,m\rangle
\langle j,m|e^{iJ_y\phi_d}J_y|j,q'\rangle\langle j,q'|\psi_{\rm in}\rangle\nonumber\\
&=4\sum_j\sum_{q,q'=-j}^j
\langle\psi_{\rm in}|j,q\rangle
\langle j,q|J_y^2|j,q'\rangle
\langle j,q'|\psi_{\rm in}\rangle\nonumber\\\,.
\label{fisher}
\end{align}
Since $J_y$ does not change the total angular momentum, we can write
this as
\begin{align}
F(\phi_d)
=4\sum_{j,j'}\sum_{q,q'=-j}^j
\langle\psi_{\rm in}|j,q\rangle
\langle j,q|J_y^2|j',q'\rangle
\langle j',q'|\psi_{\rm in}\rangle
=\langle\psi_{\rm in}|K^2|\psi_{\rm in}\rangle\,.
\label{eq:Fone}
\end{align}
For an input state that has real expansion coefficients in the number
basis, we also have
\begin{align}
\langle\psi_{\rm in}|K|\psi_{\rm in}\rangle
&=2\sum_{j,m}\langle\psi_{\rm in}|J_y|j,m\rangle\langle j,m|\psi_{\rm in}\rangle\nonumber\\
&=-i\sum_{j,m}\langle\psi_{\rm in}|(J_+-J_-)|j,m\rangle\langle j,m|\psi_{\rm in}\rangle\nonumber\\
&=-i\sum_{j,m}[C_+(j,m)\langle\psi_{\rm in}|j,m+1\rangle-C_-(j,m)\langle\psi_{\rm in}|j,m-1\rangle]
\langle j,m|\psi_{\rm in}\rangle\nonumber\\
&=-i\sum_{j,m}[C_+(j,m)-C_-(j,m+1)]
\langle\psi_{\rm in}|j,m+1\rangle\langle j,m|\psi_{\rm in}\rangle\nonumber\\
&=0\;,
\label{eq:zero}
\end{align}
where $C_{\pm}(j,m)=\sqrt{j(j+1)-m(m\pm1)}$.  The reality of the
expansion coefficients is used to get to the last expression, which
vanishes because $C_+(j,m)=C_-(j,m+1)$.  The identity~(\ref{eq:zero})
allows us to write the classical Fisher information~(\ref{eq:Fone})
as
\begin{align}
F(\phi_d)
=\langle\psi_{\rm in}|K^2|\psi_{\rm in}\rangle
-(\langle\psi_{\rm in}|K|\psi_{\rm in}\rangle)^2=\mathcal{F}_{dd}\,,
\label{eq:Ftwo}
\end{align}
where $\mathcal{F}_{dd}$ is the quantum Fisher information for the
differential phase shift [see Eq.~(8) of the text].

The equality of the classical and quantum Fisher informations applies
to all input states that have real expansion coefficients in the
number basis.  It also applies to states obtained from such input
states by rotating both input modes by the same angle $\theta$, i.e.,
by applying $e^{iN_s\theta}$ to both modes.  Applying this rotation
to the state~(\ref{eq:alphaS0}) removes the assumption that $\alpha$
is real.

\subsection{Modal entanglement after the initial beam splitter}

Here we present simple argument, due to Z.~Jiang, to show that only coherent states in the secondary input port yield a product state after the initial beam splitter.

We begin by noting that the product input state can be written as
\begin{align}
|\psi_{\rm in}\rangle=|\alpha\rangle\otimes|\chi\rangle=D(a_1,\alpha)|0\rangle\otimes D(a_2,\beta)|\chi_0\rangle\,,
\end{align}
where $D(a,\alpha)=e^{\alpha a^\dagger-\alpha^*a}$ is the displacement operator, $\beta=\langle \chi_0|a_2|\chi_0\rangle$ is the mean amplitude of the state $|\chi\rangle$, and $|\chi_0\rangle=D(a_2,-\beta)|\chi\rangle$ has this mean amplitude removed.  The state after the initial
beamsplitter is
\begin{align}
B|\psi_{\rm in}\rangle
=D\Bigl(a_1,(\alpha-i\beta)/\sqrt2\Bigr)\otimes D\Bigl(a_2,(\beta-i\alpha)/\sqrt2\Bigr)
B|0,\chi_0\rangle\,.
\end{align}
The two displacement operators act locally in the two arms, so the modal entanglement of $B|\psi_{\rm in}\rangle$ is the same as the modal entanglement of $B|0,\chi_0\rangle$.  Specifically, the displacement of the primary mode before the beamsplitter does not contribute to the entanglement after the beamsplitter~\cite{Asboth2005a}; likewise, the displacement of the secondary mode before the beamsplitter does not contribute to the post-beamsplitter entanglement. Showing that $B|\psi_{\rm in}\rangle$ is a product state only if $|\chi\rangle$ is a coherent state is equivalent to showing that $B|0,\chi_0\rangle$ is a product state only if $|\chi_0\rangle$ is the vacuum state.

If $B|0,\chi_0\rangle$ is a product state, then after the beamsplitter,
\begin{align}
0=\langle a_1^\dagger\rangle\langle a_2\rangle=\langle a_1^\dagger a_2\rangle=\langle0,\chi_0|B^\dagger a_1^\dagger a_2B|0,\chi_0\rangle\,.
\end{align}
Using $B^{\dagger} a_1^{\dagger} a_2 B = \frac{1}{2}(-ia_1^{\dagger} a_1+ ia_2^{\dagger}a_2+a_1^{\dagger}a_2+a_2^\dagger a_1)$, we get
\begin{align}
0=\frac{i}{2}\langle\chi_0|a_2^\dagger a_2|\chi_0\rangle\,.
\end{align}
showing that $|\chi_0\rangle$ is the vacuum state, as promised.

We have shown that any state other than a coherent state into the secondary input port leads to modal entanglement after the beamsplitter.

\end{widetext}

\end{document}